
\input harvmac
\Title{TOHO-FP-9242}{Metric from Non-Metric Action of Gravity}
\vglue 1cm
\centerline{Kiyoshi KAMIMURA {$^\dagger$},~ Shinobu MAKITA }
\centerline{Department of Physics, Toho University}
\centerline{Funabashi, Japan 274}
\centerline{and}
\centerline{Takeshi FUKUYAMA
{$^\star$}}
\centerline{Department of Physics, Ritsumeikan University}
\centerline{Kyoto, Japan 603}
\vglue 2cm

The action of general relativity proposed by Capovilla, Jacobson and Dell is
written in terms of $SO(3)$ gauge fields and gives Ashtekar's constraints for
Einstein gravity. However, it does not depend on the space-time metric nor its
signature explicitly. We discuss how the space-time  metric is introduced from
algebraic relations of the constraints and the Hamiltonian by focusing our
attention on the signature factor. The system describes both Euclidian and
Lorentzian metrics depending on reality assignments of the gauge connections.
That is, Euclidian metrics arise from the real gauge fields. On the other
hand, self-duality of the gauge fields, which is well known in the Ashtekar's
formalism, is also derived in this theory from consistency condition of
Lorentzian metric. We also show that the metric so determined is equivalent
to that given by Urbantke, which is usually accepted as a definition of the
metric for this system.

--------------------------------------------------------------------------

{$^\dagger$}(KAMIMURA@JPNRIFP.BITNET)~{$^\star$}(FUKUYAMA@JPNRIFP.BITNET)

\Date{28/02/92} 

A new form of canonical gravity proposed by Ashtekar\ref
\Ash{A.Ashtekar, Phys. Rev. Lett. {\bf 57}, 2244 (1986),
Phys. Rev. {\bf D36}, 1587 (1987), "Lectures on
Non-Perturbative Canonical Gravity",(World Scientific, 1991).}~
has various nice features. It is a non abelian gauge theory with the gauge
group SO(3) (or SU(2)). The first class constraints generating
the general coordinate and the gauge transformations are of polynomial forms.
However, the canonical variables are complex valued and must satisfy so called
{\it reality conditions}.
It is formulated starting from an action which is a sum of Einstein and
complex total divergence terms and significance of the reality conditions
has been discussed\ref\FK
{K.Kamimura and T.Fukuyama, Phys. Rev. {\bf D41}, 1105 (1990),
                            Phys. Rev. {\bf D41}, 1885 (1990).}.

When we find SO(3) as a fundamental symmetry group, it is more natural
to describe the Lagrangian regarding the gauge connections as field variables.
The action in terms of SO(3) gauge fields is found by taking
a Legendre transformation backward
from the Hamiltonian to Lagrangian formalism .
It is first given by Capovilla, Jacobson and Dell(CJD)\ref\CDJ
{R.Capovilla, T.Jacobson and J.Dell, Phys. Rev. Lett. {\bf 63}, 2325 (1989).}
 and after generalized to those including cosmological constants\ref\Ben
{An extensive review is given in  I.Bengtsson, "Ashtekar's variables and
the Cosmological Constants", G\"oteborg preprint ITP91-33,(1991). }.
The CJD action for the Einstein gravity is
\eqn\ea{
{\cal L}={-1\over{4 \eta}} [G^{ab}G_{ab}-{1\over 2}G^a_{\; a}G^b_{\; b}],}
where $G_{ab}$ is the building block of the action. It is defined in terms of
 $SO(3)$ gauge field strength by
\eqn\eb{
G_{ab}={1\over 4}\epsilon^{\mu\nu\rho\sigma}F_{\mu\nu a}F_{\rho\sigma b},}
\eqn\ec{F_{\mu\nu a}=
\partial_{[\mu}A_{\nu]a}+g\epsilon_{abc}A_{\mu}^{\; b} A_{\nu}^{\; c}.}
$\eta$  in the Lagrangian is a scalar density multiplier field.
The action is invariant under general coordinate transformations
as well as $SO(3)$ gauge transformations.
A characteristic property of the action is that it does not depend on any
space-time metric. Especially there is {\it a priori} no sign of signature
of metric (Euclidian or Lorentzian). $\epsilon^{\mu \nu \rho \sigma} $
is simply a Levi-Civita symbol, $\epsilon^{0123}=1$.
Another is absence of reality conditions accompanied in the Ashtekar
formalism. It seems that they may be added {\it by hand} to have
Lorentzian gravity.

In this paper we will discuss how the space-time metric arises from the action
{}~\ea,~which does not depend on the metric but has the general coordinate
invariance. Especially we focus our attention on the signature of
space-time metric. We find that the CJD action describes both Euclidian
and Lorentzian metrics of general relativity depending on the reality
assignments of the gauge connections. The reality conditions are derived
from reality of the metric and the sign of its signature.
They are found in known forms that $A_{\mu a}$ must be real for Euclidian
metric and must be a complex self-dual spin connection for Lorentzian one.

In Hamiltonian formalism of the action  \ea ~ there appears
a set of first class constraints,
\eqn\eca{
\pi_\eta = 0, }
\eqn\ec{\pi^{0a} = 0, }
\eqn\ed{J^a = D_i \pi^{ia} = 0,}
\eqn\ee{T_j = \pi^{ka} F_{jka}=0}
and
\eqn\ef{
H_0 = {1\over 2} \epsilon_{abc} \pi^{ia} \pi^{jb} F_{ij}^{\quad c}=0,}
where $\pi_\eta$ and $\pi^{\mu a}$ are momenta conjugate to $\eta$ and
$A_{\mu a}$, respectively and $D_\mu$ is $SO(3)$ covariant derivative.
The constraints \eca~ and \ec~  tell that ~$\eta$~and~$A_{0a}$~
are arbitrary multipliers.
Equation \ed~ is a Gauss law constraint for $SO(3)$. The last two
\ee~ and \ef~are reflecting the general covariance of the Lagrangian.
The canonical Hamiltonian  ${\cal H} = p \dot q - {\cal L}$~  becomes
\eqn\eg{
{\cal H}=\pi^{0a}\dot A_{0a} -A_{0a}J^a+ {1\over 2}
         \epsilon^{ijk}E_{ia}{\bf B}_j^{\; a}T_k+{\eta\over{2detB}}H_0,}
where the electric field is $E_{ia} \equiv F_{0ia}$ and
the magnetic field is $B^{ia} \equiv {1\over 2}\epsilon^{ijk}F_{jk}^{\quad a}$.
${\bf B}_{ia}$ is its inverse and $det B (\equiv det B^{ia})$ is assumed
to be non vanishing. In~\eg~ we have used only ~
$ ( p - {{\partial L}\over{\partial \dot q}}  )^2 = 0$~as strong equality
but ~ $ ( p - {{\partial L}\over{\partial \dot q}}  )   = 0$~has never been
used. By his prescription we can obtain correct forms of
multipliers on constraints in terms of $p,q$ and $\dot q$ \ref\KK
{K.Kamimura, Nuovo Cimento {\bf 68B}, 33 (1982). }.

The space-time metric is introduced if we can identify
the constraints with diffeomorphism generators
$H_\perp$ and $H_j (j=1,2,3)$. In a metric space $H_j$'s generate
transformations of three coordinates of space-like hyper-surfaces.
$H_\perp$, on the other hand, deforms the hyper-surface in its normal
direction $n_\mu$ . They satisfy the following algebra \ref\Tb
{C.Teitelboim, Ann. of Phys. {\bf 79}, 542 (1973).}
\eqn\eh{\lbrace H_i(x),\; H_j(y)\rbrace =H_i(y){\partial\over \partial x^j}
       \delta (x-y)-H_j(x){\partial \over \partial y^i}\delta(x-y), }
\eqn\ei{\lbrace H_j (x),\; H_\perp (y)\rbrace =H_\perp (x){\partial \over
       \partial x^i}\delta(x-y) }
and
\eqn\ej{\lbrace H_\perp (x),\; H_\perp (y) \rbrace =-\epsilon(\gamma ^{ij}
(x)H_j(x)+\gamma^{ij}(y)H_j(y)){\partial \over \partial x^i}\delta(x-y)
,}
where $\epsilon = n_\mu n^\mu$ is a signature of the metric {\it i.e.}~
 $\epsilon= +1 $ ~for Euclidian signature and $-1$ for Lorentzian one.
$\gamma^{ij}$~ is the induced metric of the space-like hyper-surface
and appears in the Poisson bracket of $H_{\perp}$'s as the structure function.

$T_j$'s in \ee ~ satisfy the same algebra as that of $H_j$'s in \eh
\eqn\ek{\lbrace T_i(x),\; T_j(y) \rbrace =T_i(y){\partial \over
\partial x^j}\delta(x-y)-T_j(x){\partial \over \partial y^i}\delta(x-y)}
and they are identified without factor ambiguity
\eqn\el{T_j = H_j.}
Here and hereafter equalities are satisfied up to the $SO(3)$ gauge
constraint;~ $J^a=0$. It means that additional $SO(3)$ transformations
are associated
in the commutators. The algebra of $T_j$ and $H_0$ becomes
\eqn\em{
\lbrace T_j(x),\; H_0(y) \rbrace =H_0(x){\partial \over \partial x^j}
\delta(x-y)-H_0(y){\partial \over \partial y^j}\delta(x-y).}
In order to connect~$H_0$~ with ~$H_\perp$~, we change the density
weight of ~$H_0$~ by dividing it by $\pi^{1/2}$ ,
 $( \pi \equiv \mid det\pi^{ia} \mid ).$ The Poisson bracket
\eqn\en {
\lbrace T_j(x),\; {1\over \pi^{1/2}}H_0(y)\rbrace
 ={1\over \pi^{1/2}}H_0(x) {\partial \over \partial x^j}\delta (x-y)}
has the same form as \ei~ and the Poisson bracket corresponding to \ej~ is
\eqn\ena{
\lbrace {1\over \pi^{1/2}} H_0(x) , {1\over \pi^{1/2}} H_0(y) \rbrace =
-[{\pi^{ia}\pi^j_{\>a} \over \pi} T_j(x) +{\pi^{ia}\pi^j_{\>a} \over \pi}
T_j(y)]
{\partial \over \partial x^i}\delta(x-y).}
It must be noted that the over all constant factor $a$ of $H_\perp$ is not
fixed algebraically,
\eqn\eo{H_\perp = {a\over \pi^{1/2}}H_0.}
It is assumed that $\pi$ dose not vanish. The case of  $\pi=0$ corresponds
to degenerate metric, which is not  prohibited in the action \ea.
When $\pi$ vanishes, in a region of parameter space of $x^\mu$~,
the metric (of weight zero) is not well defined there.
The CJD action may describe such a generalization of the Einstein theory \Ben.

Using $H_\perp$ in \eo~and comparing the structure functions of
 ~ \ej ~ and ~\ena~, we find a expression of three metric $\gamma^{ij}$~ as
\eqn\ep{
\gamma^{ij}=\epsilon a^2{\pi^{ia}\pi^j_{\; a} \over \pi}.
}
The remaining components of the metric are determined by examining
a generator of surface deformation of amount $\delta y^\mu$ \Tb
\eqn\eqq{G[\delta y^\mu]=\int \delta y^{\perp}H_{\perp} +\delta y^jH_j.}
In Hamiltonian formalism the space-like surface is labeled by $x^0$ itself
and the Hamiltonian is the generator of $x^0$ translation. In this case
\eqn\er {\delta y^\perp = \epsilon n_\mu \delta y^\mu =
{\epsilon \over{(\epsilon g^{00})^{1\over2}}}\delta x^0,\qquad
\delta y^j = \gamma^{jk} y^\mu_{\; ,k} g_{\mu \nu} \delta y^\nu
= -{ g^{0j} \over g^{00}}\delta x^0 }
and
\eqn\eu{\gamma^{ij}=g^{ij}-{{g^{0i}g^{0j}}\over{g^{00}}}.}
By comparing the Hamiltonian \eg~ with \eqq~ and \er~we find
\eqn\et{
\alpha \equiv{1\over{\sqrt{\epsilon g^{00}}}} =
{{\epsilon \eta  \pi^{1\over2}}\over{2 a\; detB}}}
and
\eqn\es{{\beta}^k \equiv -{g^{0k}\over g^{00}} =
{1\over2} \epsilon^{ijk} E_{ia}{\bf B}_j^{\;a}.}

We will examine whether the metric $g^{\mu\nu}$ are determined from \ep,~
\eu,~\et~ and \es~ consistently. The requirements for
the metric are  the following:~
{}~1) all the components are real quantities and
{}~2) $\epsilon g^{00}$ must be positive.
{}~3) $\gamma^{ij}$ is positive definite so that the equal-time space
spans a space-like hyper-surface.

The lapse function $\alpha$ in \et~is a multiplier of {\it secondary}
first class constraint $H_\perp$~ and is taken to be positive as a
result of gauge fixing condition on $\eta$.
The shift vector ${\beta}^j$ in \es ~is a multiplier of {\it primary}
first class constraint $T_j$~and is taken to be real as the gauge choice.
It means that for any  $\beta^j$~ chosen  in the Hamiltonian  the equation
{}~\es~ is always satisfied as a result of the equations of motion.
$\gamma^{ij}$, on the other hand, is determined dynamically by \ep.
Assuming $\pi^{ia}$'s are real variables, $\epsilon a^2$ is required to be
positive ;
\eqn\eqv{\epsilon = sign (a^2 ).}
The constant  $a$ is not determined from the $H_\perp$ algebra \ei~ and \ej.
However it enters in the Hamiltonian through $H_\perp$ and complex choices
of ~$a$~ generally violate reality property of the canonical variables.
For real values of ~$ a \quad (\equiv {1\over {\sqrt\kappa}}),~\epsilon$~is
$+1$ (Euclidian) and all quantities including the gauge connection $A_{ia}$'s
are taken to be real variables.
That is, the action \ea ~describes metrics with Euclidian signature
when the dynamical variables are real fields.
It is consistent with the observation that the Ashtekar connections are
real variables in the  Euclidian signature.

In order to have Lorentzian signature ~$\epsilon = - 1 , \;a\;$ must be a
pure imaginary number (say $a \equiv {i\over{\sqrt\kappa}}). $ ~
In this case $A_{ia}$'s cannot be real variables since the reality of
$\pi^{ia}$ and $A_{ia}$ are not preserved during time development.
We can show the following reality conditions on the canonical variables,
\eqn\ew{
\pi^{ia \dagger}=\pi^{ia},}
\eqn\ex{
(igA_{ia})^ \dagger=(igA_{ia})+i\epsilon_{abc}\omega_i^{\;bc}(\pi^{ja})}
are consistent with the Lorentzian signature, when  $\alpha$~ and ${\beta}^j $~
 are real and the multiplier $A_{0a}$ of the Gauss law constraint satisfies
\eqn\exo{
A_{0a}^ \dagger=A_{0a}-{2\over{ig}}\omega_{00a}(\pi^{ia},\alpha,\beta^j).}
Here $\omega_\mu^{\; AB}$'s are components of spin connection which
will be defined later.
The reality conditions \ew~ and \ex~ are same as ones imposed in Ashtekar's
formalism \Ash ~while \exo~is the reality condition on the multipliers
of the first class  constraints \FK.

We will see how the reality conditions come out.
We introduce spin connections  $\omega_{\mu AB}(A=0,1,2,3)$~ as auxiliary
functions. They are defined using a torsion free condition
\eqn\eac{
\partial_{[\mu}e_{\nu]A} + \omega_{[\mu A}^{\;\quad B} e_{\nu ] B}=0}
with the tetrad variables $e_{\mu a}$
\eqn\ead{
        e_{ia}={\pi_{ia} \over{ (\kappa \pi)^{1/2}}},\quad e_{00}=\alpha,\quad
        e_{0a}=\beta^je_{ja},\quad e_{i0}=0.}
Here $\pi_{ia}=\gamma_{ij}\pi^j_{\;a}$~and $ \gamma_{ij}$ and $e_{ia}$
are the inverse of $\gamma^{ij}$ and $e^{ia}$, respectively.
{}From the torsion free condition the components of the spin connection are
solved as functions of $\pi^{ia},\; \alpha,\; \beta^j$ and ${\dot\pi}^{ia}$.
Among those, $\omega_{iab}$ is given as a  function of $\pi^{ia}$'s only,
\eqn\eiab{
   \omega_{iab}=-{1\over 2}e_i^{\;c}(A_{abc}-A_{bca}-A_{cab});
\qquad A_{abc} \equiv e^j_{\; a}e^k_{\; b} \partial_{[j}e_{k]c}.}
Other components, $\omega_{i0a},\; \omega_{0ab}$~ and $\omega_{00a}$,~
depend on the time derivative of $\pi^{ia}$'s. ${\dot\pi}^{ia}$ is evaluated
 using Hamilton's equation of motion with the Hamiltonian \eg~
\eqn\egz{
{\cal H}=\pi^{0a}\dot A_{0a} -A_{0a}J^a+ {\beta^j}T_k+
         \epsilon \alpha H_{\perp}.}
In the last term, ~$\epsilon a~={1\over{I\sqrt\kappa}}$~
and the factor ~$I$~ is equal to 1 for $\epsilon =1$~ and ~$i$~ for
$\epsilon =-1$~. The Hamilton's equation for $\pi^{ia}$ is
\eqn\eaa {\dot\pi^{ia}=
D_j[{\alpha\over {I(\kappa \pi)^{1\over2}}}\epsilon^{abc}\pi^j_{\;b}
\pi^i_{\;c}]+D_j[\beta^j\pi^{ia}-\beta^i\pi^{ja}]
+g\epsilon^{abc}\pi^i_{\;b}A_{0c}.}
Thus the components of the spin connection are given as functions of
{}~$\pi^{ia},~ \alpha,~ \beta^j$~ and ~$A_{\mu a}$~
(and their spatial derivatives);
\eqn\eooa{
\omega_{00a}=e^i_{\;a}[\partial_i\alpha+(IgA_{ib}
+{I\over 2}\epsilon_{bcd}\omega_i^{\;cd})\beta^j e_j^{\;b}],
}
\eqn\eoab{
\omega_{0ab}={1\over I}\epsilon_{abc}\omega_{00}^{\quad c}
             -g\epsilon_{abc}A_0^{\;c}}
and
\eqn\eioa{
\omega_{i0a}={I\over 2}\epsilon_{abc}\omega_i^{\; bc}
             -IgA_{ia}.}
Equations \eiab~and \eooa-\eioa~ are expressions of the spin connections
in terms of the canonical variables and multipliers $\alpha,~\beta^j$~
and $A_{0a}$.

Now we return to the reality conditions.
The reality of $\pi^{ia}$ requires that of ${\dot\pi}^{ia}$ also.
It turns out that all components of the spin connection,
which are defined from
the torsion free condition \eac, are real quantities. The r.h.s. of eq.\eiab~
is apparently real while those of \eooa-\eioa~ are not.
The reality condition of
$\pi^{ia}$ actually requires that they must be real quantities.
It turns to be reality conditions on the gauge connections.
Note \eoab~ and \eioa~ are expressed as covariant form
\eqn\eia{
 {Ig \over 2}A_{\mu a}={1\over 2}(\omega_{\mu 0a}-{I\over 2}\epsilon
_{abc}\omega_\mu^{\;bc})\; \equiv \omega_{\mu 0a}^{(+)}}
with real components of spin connection. For the Euclidian metric,
{}~$\epsilon=+1$~ and ~$I=1$.~ It says that $A_{ia}$~'s are real
canonical variables and $A_{0a}$'s are real multipliers.
For the Lorentzian metric, $\epsilon =-1$~ and $I=i$,~
$A_{ia}$~'s are canonical variables satisfying the reality condition \ex~
and multipliers $A_{0a}$'s must satisfy the reality condition \exo.
The latter condition tells that $A_{0a}$'s are not chosen arbitrarily but the
imaginary parts are related to $\alpha$~ and $\beta^j$.

    The equation \eia~ tells that $A_{\mu a}$'s are self-dual spin connections
with respect to the internal indices both for $\epsilon = \pm 1$.
By this assignment of reality conditions the total Hamiltonian density \egz~
is shown to be real. It guarantees that the reality condition \ex~ holds also
at any time when it is satisfied initially. That is, time derivative of
$A_{ia}^\dagger$~ is equal to complex conjugate of~ $\lbrace A_{ia},
{\cal H}\rbrace$. The Einstein equations follow from the equation of motion
for $A_{ia}$ when it is combined with the constraints \ee~ and \ef.

The metric determined above from the algebraic relations of generators
is equivalent to that given by
Urbantke \ref\Urban{H.Urbantke, J. Math. Phys. {\bf 25},2321 (1984).}.
The latter is a symmetric contravariant tensor composed of $SO(3)$ gauge
fields  in a gauge invariant form,
\eqn\eur{
g^{\mu\nu}={\; A \; \over 3!}
\epsilon^{\mu\alpha\beta\gamma}\epsilon^{\nu\rho\sigma\lambda}F_{\alpha\beta a}
F_{\rho\sigma b}F_{\gamma\lambda c}\epsilon^{abc},}
where $A$ is a scalar density of weight $-2$ and will be determined below.
In terms of electric and magnetic fields they  are
\eqn\eura{
g^{00}={4\; A \; det \;B},}
\eqn\eurb{g^{0k}=-{g^{00}\over 2} \epsilon^{ijk} E_{ia}{\bf B}_j^{\;a}}
and
\eqn\eurc{g^{ij}={g^{00} \over 8}\epsilon^{ikl}\epsilon^{jmn}
(E_{(ka}{\bf B}_{m)}^{\;a})(E_{(lb}{\bf B}_{n)}^{\;b})
+{{g^{0i}g^{0j}}\over g^{00}}.}
$A$ is fixed by comparing \eura~ with ~\et~ as
\eqn\eurd{A={{(\epsilon a^2)\;det B}\over{\eta^2\;\pi}}.}
The second equation \eurb~ is exactly same as \es.
In order to show the third one \eurc~ gives the same three metric
$\gamma^{ij}$ as in \ep~
we must write the first term in terms of canonical variables.
It is indeed performed and \eurc~ gives same metric of \ep.
Although the metric ~\eur~ contains momentum variable $\pi$.  It
is written in terms of Lagrangian variables only in a following form
\eqn\eure{
{\sqrt{\epsilon g}}g^{\mu\nu}={I\kappa \over 12\; \eta}
\epsilon^{\mu\alpha\beta\gamma}\epsilon^{\nu\rho\sigma\lambda}F_{\alpha\beta a}
F_{\rho\sigma b}F_{\gamma\lambda c}\epsilon^{abc},}
where $g=det{g_{\mu\nu}}$. Here again the same reality conditions
must be assigned to obtain physically acceptable metrics.

In summary we have shown that the CJD action describes both Euclidian and
Lorentzian metrics of general relativity depending on the reality
assignments of the gauge connections.
That is, $A_{ia}$ must be real for Euclidian metric $(\epsilon=I=1)$ and
must satisfy the reality condition \ex~ for Lorentzian one
$(\epsilon=-1,\;I=i)$. These reality conditions may be required on the initial
data of the canonical variables in  classical theory and  on the additional
measure factor in the inner products in  quantum theory\FK.
The feature of CJD action is not only to give both Euclidian and Lorentzian
metrics but also degenerate one\CDJ\Ben. They may play important roles in
quantum gravity and quantum cosmology.
\listrefs
\end